# On the Aspect of Plane of Appearance of Jahn-Teller and Renner-Teller Intersections in Tetra Atomic System – A Case Study with HCNO$^+$

**Rintu Mondal and Debasis Mukhopadhyay***

Department of Chemistry, University of Calcutta, Kolkata, India.

## Abstract

Search for configuration space with well-defined topological (Berry) phases corresponding to Jahn-Teller (JT) conical intersection (CI) and Renner-Teller(RT) parabolic intersection (PI) in the linear tetra-atomic molecular system on introduction of bending, reveal the interesting aspect that these potential intersections may appear in molecular plane as well as out of the molecular plane. While understanding this aspect is important for following the class of phenomena led by potential intersections, till date studies on molecular systems including pairs like ($C_2H_2^+$, HCNH) as well as ($N_2H_2^+$, HBNH$^+$), have not been able to clarify the issue. The present paper embodies calculation of non-adiabatic coupling terms (NACT's) involving four low-lying states of slightly bent HCNO$^+$, a motivated choice of tetra-atomic with all four different atoms, to study this aspect associated with JT-CI and RT-PI in slightly bent linear system. The plane of appearance of these effects, has been advocated to be related to electronic configuration of the concerned states of the molecular system.

*e-mail : dm.chem.cu@gmail.com

# 1. Introduction:

Energetics of low-lyling electronic states of small molecules with various types of curve-crossings have been conveniently taken as an avenue to study the electronic non-adiabatic effects on the adiabatic surfaces, as obtained with Born-Oppenheimer approximation[1,2]. In general, a molecule is a system having the fast moving electrons, driven by a slow moving external field due to vibrating or rotating nuclei and thus the Longuet-Higgins (LH) phase[3] acquired by the adiabatic electronic eigenfunctions on surrounding a point of degeneracy in a configuration space (CS) has been identified as the topological (Berry) phase[4] in the molecular systems. It is to note that the non-adiabatic coupling terms (NACT's) between adiabatic electronic states are responsible for deviation from, as well as for breakdown of B.O. approximation. Hence it was felt necessary to go to a diabatic framework through adiabatic to diabatic transformation (ADT). In this regard though the works of Hobey and Mclachlan[5], Smith[6] are historically important, in what follows we would follow the theoretical framework, first introduced by Baer[7] and later championed by Baer with collaborators[8-20] and several other groups[21-28].

It is now established[13] that ADT-angle is conceptually related to the NACT's between different interacting electronic states. In quantum chemical calculations involving a number of strongly interacting electronic states of a molecule forming a Hilbert sub-space (HSS), adiabatic to diabatic transformation angle (ADT-angle or mixing angle ) may be identified[11] with the topological (Berry) phase of the molecule. While for a study in detail we refer to Refs 13 and 14, for completeness of the paper we mention the following. Since the NACT becomes singular at the point of degeneracy, the corresponding ADT matrix may become multi-valued, which is again related to the multi-valued-ness of the electronic eigenfunctions. The nuclear Schrödinger equation cannot be solved for a multi-valued potential. In order to have a physical meaning newly constructed diabatic potential has to be single-valued; this is guaranteed only when the ADT matrix at the end of the closed contour (i.e., the D-matrix) is diagonal having a value of ±1. The diagonal D-matrix is associated with the ADT angle (or mixing angle) at the end of the closed contour as a multiple of π(or zero). NACT matrix (NACM) yielding diagonal D-matrix in a HSS is referred to as being quantized in the corresponding configuration space.

In this context it is worth mentioning that a closed contour, associated with a topological phase of π, surrounds a single Jahn-Teller conical intersections (JT-CI)[29] while the one associated with a topological phase of 2π, surrounds a single Renner-Teller parabolic intersection( RT-PI)[30]. Moreover, such designation is in commensuration with the terminologies of intersections according to the symmetry of the states involved. In practical calculations, the topological phases may deviate from the physically significant values, due to the inadequacy in description of HSS, where the states outside the HSS chosen, interact with the states within.

The current paper, while presents a similar study on a linear molecule HCNO$^+$, the motivation is to resolve the issue of the plane of appearance of JT-CI's and RT-PI's involving states evolved from degenerate X$^2\Pi$ states, on introduction of bending. In order to convey the purpose of writing this paper we present a brief review, in what follows. For a tetra atomic system one atom may be chosen as the probe to define the closed contour. The first of such studies was done by Halasz etal[17] to explore the interplay of JT and RT effect in slightly bent linear tetra-atomic C$_2$H$_2$$^+$. In this study, two different cases (Fig.1 of Ref. 17a) have been considered : (a) one terminal atom, shifted from the molecular axis, surrounds the molecular linear axis. (b) one of the terminal atoms shifted from the molecular axis, is clamped at a distance, while another terminal atom, shifted from the molecular axis, is allowed to surround the molecular axis in a plane (plane I), perpendicular to the plane (plane II ) formed by the rest three atoms. This defines a circular contour with its center on the molecular axis. While the angle φ between the plane I and plane II is zero(2π) or π, four atoms are coplanar and the plane is designated as molecular plane. For φ ≠ zero(2π),π the four atoms are not co-planar and this corresponds to non-planar configuration of the molecule; later we have described this situation by writing that the probe atom is out of molecular plane. While one linear system is originally RT, as revealed[17] in study of case (a), Halsaz etal[17] demonstrated the onset of JT intersection in case (b). Observation of such changes made the field interesting and explorations[18,24-28] were continued to study the generality as well as different aspects related to this phenomena. It is important here to point out that, in case (b), the molecular configuration, in general, may be characterized only by C$_1$ point group and thus the electronic states can no longer be characterized by spatial symmetry. Again for dihedral angle, φ = zero(2π),π, the molecular configuration is planar and attains C$_S$ symmetry. Thus for planar

configuration ($\varphi$ = zero($2\pi$),$\pi$), states are tagged by symmetries A´ and A″, while for non-planar configuration ($\varphi \neq$ zero($2\pi$),$\pi$) states are tagged only by symmetry A. In an extensive study of linear system HCNH, Das etal[24,25] demonstrated that the JT-CI between the two lowest states originated from collinear $X^2\Pi$ state on introduction of bending, appear only in some non-planar configuration (i.e., out of molecular plane), in contrast to the appearance of JT-CI between such two lowest states of slightly bent $C_2H_2^+$ in molecular plane. Consequently the question has been raised[25,28] why HCNH and $C_2H_2^+$ have been different in this aspect of plane of appearance of JT-CI between the two states originated from $X^2\Pi$ state on introduction of bending. We mention here the conjecture[25] proposing that the reason may lie in the fact that HCNH is different in possessing a collinear axis made up from two hetero atoms (C,N) differing in electronegativity. We like to point out that Bene etal[40] tried to justify the specialty of the (1,2) CI in HCNH by considering the non-symmetric nature of the molecule in comparison to $C_2H_2^+$ and commented that more work is needed for a conclusive understanding. Das etal[28] have recently reported different result for similar calculations involving a similar pair of linear molecular systems $N_2H_2^+$ and $HBNH^+$. Similar to the pair $C_2H_2^+$ and HCNH, here $HBNH^+$ has a collinear axis made up of two hetero atoms (B,N), differing in electronegativity in contrast to $N_2H_2^+$. They demonstrated that while as to RT effect there has been a regular observation, JT-CI involving two lowest states (originating from $X^2\Pi$ state on introduction of bending) appear in molecular plane for $HBNH^+$ and out of plane for $N_2H_2^+$. In summary, for states evolved from linear $X^2\Pi$ state on introduction of bending, we highlight the following result from the literature : for molecular systems $C_2H_2^{+,17}$ and $HBNH^{+,28}$, JT-CI between two lowest states evolved, appear in molecular plane, while for HCNH[25] and $N_2H_2^{+,28}$, that appear out of molecular plane. Hence argument was placed[28] that the symmetric/non-symmetric nature of the molecules is not related to the issue of plane of appearance of JT-CI between two states, evolved from $X^2\Pi$ state.

In the current paper we report a study on $HCNO^+$, which is similar to the earlier studies[17,18,24,25,28] in methodology but with an analysis **distinct in contribution to the understanding of the issue explained above**. The electronic structure of $HCNO^+$ is special in the respect that on introduction of bending, each of the two low-lying linear $X^2\Pi$ states splits into two states. The protocol that introduced by Halasz etal[17] and later widely used in several works[18,24,25,28] has

again been used to identify intersections with calculation of NACTs, ADT angles and topological phases; but **the analysis addresses the question mentioned above, which remains unsolved till date**. We have demonstrated that the lowest pair of states, evolved from the lowest linear $1^2\Pi$ state, shows a JT-CI in molecular plane, while the upper pair of states, evolved from the upper linear $2^2\Pi$ state, shows RT intersection out of molecular plane. We have classified the linear $X^2\Pi$ states in $HCNO^+$ as well as in the earlier studied molecules $C_2H_2^+$, HCNH, $N_2H_2^+$ and $HBNH^+$ in order to reveal a correlation between the plane of appearance of potential intersections (JT-CI & RT-CI) between the states evolved from these linear states and the nature of molecular orbitals ($\Pi/\Pi^*$) to which the linear states may be tagged by origin.

It is worth to point out here that 'Beyond B.O.' treatment of Adhikari etal[31-34], work of Varandas[35] related to the LH theorem[3] as well as generalized B.O. theory by the group of Varandas[36,37] have also been successful in non-adiabatic treatment involving low-lying electronic states. But the issue of plane of appearance of potential intersections has not been addressed in these works.

In the theoretical framework section (Section 2) we briefly describe the theoretical background for the completeness of the paper. The section on numerical investigation and discussion (Section 3) describes the computational investigation and result in detail. In Section 4, the findings in investigation have been analyzed with the classification of the $X^2\Pi$ states involved in different linear molecules leading to the generalization of present as well as related earlier result of our group and others. In conclusion we summarize the updated status of the understanding about the plane of appearance of potential intersections between the states evolved from $X^2\Pi$ states and indicates the importance of such understanding.

## 2. Theoretical framework:

We start with the nuclear Schrödinger equation for an n-state electronic HSS, written[13,38] compactly in matrix form as

$$-\frac{1}{2m}(\nabla + \boldsymbol{\tau})^2 \Psi + (\mathbf{u} - \mathbf{E})\Psi = \mathbf{0} \tag{1}$$

Here ψ is a column vector consisting of the nuclear functions $\{\psi_j(s), j=1,..n\}$ corresponding to the Born-Oppenheimer expansion[1] for the complete wave function $\Psi(s_e,s)$

$$|\Psi(s_e,s)\rangle = \sum_{j=1}^{n}|\zeta_j(s_e,s)\rangle\psi_j(s) \quad (2)$$

$|\zeta_j(s_e,s)\rangle, j=1,..n$ are the adiabatic electronic eigen functions of the electronic Hamiltonian. Here $s_e$ and s represent electronic and nuclear coordinates respectively. **u** is a diagonal matrix containing the adiabatic potentials and **E** is a diagonal matrix consisting of the total energy corresponding to the total wave function. 'm', denoting the mass of the system is also a diagonal matrix. **τ** is the non-adiabatic coupling (vector) matrix consisting of the nuclear coordinate dependent NACT's.

At the point of intersection of two adiabatic states, the eigenvalues being degenerate, the corresponding NACT becomes singular function of nuclear coordinates and this situation poses problem to treat the equation (1). In order to get rid of **τ** matrix one may go to a diabatic framework through adiabatic to diabatic transformation. One aims to find an ADT matrix **A** such that

$$\psi = A\psi^d \quad (3)$$

$\psi^d$ is the diabatic vector consisting of the nuclear functions $\{\psi_j^d(s)\}$. **A** is an orthogonal unitary matrix of coordinates such that **τ** matrix does not appear in the Schrödinger equation for $\psi^d$,

$$-\frac{1}{2m}\nabla^2\psi^d + (\mathbf{w}-\mathbf{E})\psi^d = 0 \quad (4)$$

with **w** as the corresponding diabatic potential matrix, defined as,

$$\mathbf{w} = \mathbf{A}^\dagger \mathbf{u}\mathbf{A} \quad (5)$$

The ADT-matix **A** satisfies the differential equation[7]

$$\nabla \mathbf{A} + \boldsymbol{\tau}\mathbf{A} = 0 \quad (6)$$

or equivalently the integral equation[7,10] along a contour

$$\mathbf{A}(s|s_0, \Gamma) = \mathbf{A}(s_0|\Gamma) - \int_{s_0}^{s} ds'.\, \boldsymbol{\tau}(s', s_0|\Gamma)\, \mathbf{A}(s', s_0|\Gamma) \tag{7}$$

With the following significance of the terms involved:

$\Gamma$ is a contour to the multi-dimensional CS.

While $s_0$ and s are points on this contour, $s_0$ is the initial point of integration.

ds' is a differential vector along this contour.

Solution[7,10] of equation 7, involving the path-ordering operator $\wp$ is given as

$$\mathbf{A}(s|s_0, \Gamma) = \wp \exp\left(-\int_{s_0}^{s} ds'.\, \boldsymbol{\tau}(s'|\Gamma)\right) \mathbf{A}(s_0|\Gamma) \tag{8}$$

Here, $\mathbf{A}(s_0|\Gamma)$, the initial value of $\mathbf{A}(s)$ on contour $\Gamma$, is taken as unit matrix.

Hence, for a closed-contour, the topological matrix $\mathbf{D}$ is defined[12] as

$$\mathbf{D} = \mathbf{A}(s_0|s_0, \Gamma) = \wp \exp\left(-\oint_{\Gamma} ds.\, \boldsymbol{\tau}(s|\Gamma)\right) \tag{9}$$

Thus from equation 8,

$$\mathbf{A}(s=s_0|s_0) = \mathbf{D}\mathbf{A}(s_0) \tag{10}$$

It is pertinent here to comment that the necessary conditions for the **A**-matrix to yield single-valued diabatic potential is that **D**-matrix for a closed contour in that CS must be diagonal with norm 1. For real eigenvalues, the conclusion[12,14,20] is

$$\mathbf{D}_{ij}(\Gamma) = \pm \delta_{ij} \quad i,j = \{1,n\} \tag{11}$$

Due to the presence of degeneracy electronic adiabatic manifold may be multi-valued, while the diabatic manifold, being independent of nuclear coordinates, is single valued. As is clear from equation 10, there is a sign-flipping in the **A**-matrix and the position of (-1)s have to correspond with the electronic eigenfunctions that flip their sign. Thus, the multivalued adiabatic electronic manifold create topological effect.

In what follows, closed contours are taken as circular. Hence with

$$(\mathbf{\tau}_\varphi)_{jk} = \left\langle \zeta_j \left| \frac{\partial}{\partial \varphi} \right| \zeta_k \right\rangle, \quad \zeta_j \text{ is } j^{th} \text{ eigen state of the electronic Hamiltonian.} \tag{12}$$

$$\mathbf{A}(\varphi|q,\Gamma) = \wp \exp\left(-\int_0^\varphi d\varphi' . \mathbf{\tau}_\varphi(\varphi'|q,\Gamma)\right) \tag{13}$$

$$\text{and} \quad \mathbf{D}(q,\Gamma) = \wp \exp\left(-\int_0^{2\pi} d\varphi . \mathbf{\tau}_\varphi(\varphi|q,\Gamma)\right) \tag{14}$$

It is to note that any (2X2) orthogonal matrix may be written as

$$\mathbf{A}^{(2)}(\varphi,q) = \begin{bmatrix} \cos(\gamma_{12}(\varphi,q)) & \sin(\gamma_{12}(\varphi,q)) \\ -\sin(\gamma_{12}(\varphi,q)) & \cos(\gamma_{12}(\varphi,q)) \end{bmatrix} \tag{15}$$

Where $\gamma_{12}(\varphi,q)$, the ADT angle or mixing angle is expressed as

$$\gamma_{12}(\varphi,q) = \int_0^\varphi (\tau_\varphi)_{12}(\varphi',q)d\varphi' \tag{16}$$

For a 2-state HSS, the equations (13) and (14) are simplified.

Topological phase (Berry phase) $\alpha_{12}(q)$ is given as

$$\alpha_{12}(q) = \int_0^{2\pi} (\tau_\varphi)_{12}(\varphi',q)d\varphi' \tag{17}$$

Following equation (15), corresponding D-matrix is given by replacing $\gamma_{12}(\varphi,q)$ with $\alpha_{12}(q)$.

For an n-state HSS, with n>2 the n-dimensional ADT matrix $\mathbf{A}^{(n)}$ is expressed[9] as a product of elementary rotation matrices. In the present work we consider HSS with n=4. For 4-state HSS, the methodology to determine $\gamma_{ij}(\varphi,q)$ is involved and we refer reader to look into references 24 and 25 for this.

For completeness, we mention that in the cylindrical coordinate S=(q,φ,z) with z as the coordinate along molecular axis, q is the radius and φ is the angle associated with the rotating probe atom. The angular component of NACT should be given as $\frac{1}{q}(\tau_\varphi)_{jk}$. For simplicity, henceforth we drop the suffix φ from $\tau_\varphi$.

## 3. Numerical Investigation and Discussion :

### 3A. Details of computational study:

At the beginning of this section we point out that for a tetra-atomic linear molecular system, while one terminal atom is clamped at a position shifted from its molecular axis and the other terminal atom, shifted from the molecular axis, is rotated around the molecular axis in a plane perpendicular to the molecular axis, potential intersections are observed in molecular plane as well as out of molecular plane. We have noted, as described in introduction, while the JT-CI between two lowest states of HCNH[25] under such situation, appears out of molecular plane, that between two lowest states of $C_2H_2^+$ under similar situation, appears in molecular plane. Moreover literature[28] shows that two lowest states of the pair of molecular systems $N_2H_2^+$ and $HBNH^+$ under similar situation, show reversal of avenue for potential intersections. In the present paper we want to understand this issue of plane of appearance of potential intersections. For this purpose we have undertaken the numerical investigation of potential intersection on introduction of bending to the linear molecular system $HCNO^+$ as well as associated topological effects. The linear system, $HCNO^+$ molecule, is characterized by the C-N distance : $R_{CN}$=1.175 Å, the N-O distance : $R_{NO}$=1.198 Å and the C-H distance $R_{CH}$=1.058 Å. For our calculations we have clamped the terminal O-atom at a distance $q_O$ shifted from the molecular C-N axis, while the other terminal atom, H-atom, shifted at a distance $q_H$ from the molecular C-N axis, is rotated in a plane perpendicular to the molecular axis. Thus with the terminal H-atom taken as the probe the following types of result are considered :-

(a) Adiabatic potential energy curves for low-lying electronic states, (b) angular NACT's, $\tau_{ij}$, between $i^{th}$ and $j^{th}$ electronic states, (c) suitable mixing angles $\gamma_{ij}$ between $i^{th}$ and $j^{th}$ electronic states and the corresponding topological phases $\alpha_{ij}$ and (d) diagonal elements of ADT-matrix (A-matrix) and D-matrix. The calculation of adiabatic potential energy curves and the angular NACT's $\tau_{ij}$ between $i^{th}$ and $j^{th}$ electronic states have been performed with the quantum chemistry package of programs, MOLPRO[39], at the state-average complete active space self consistent field (SA-CASSCF) level followed by multireference configuration interaction (MRCI) using 6-311G** basis set for all the atoms. HCNO$^+$ has 21 electron in total, of which 6 electrons are taken as core and 15 electrons as valence. In SA-CASSSCF, the active space is chosen by distributing 15 valence electrons in 12 orbitals. The number of states to form an HSS have been taken, judicially on the basis of mutual proximity of the potential energy curves in the CS chosen. In calculation of NACT-elements, $\tau_{ij}$, in most of the cases DDR-technique, as prescribed in MOLPRO[39] has been undertaken. Once the NACT-matrix has been obtained for a HSS in a CS , further calculation of $\gamma_{ij}$ and diagonal elements of A-matrix and D-matrix have been performed following the protocol described in the theoretical framework section.

In what follows, we discuss on potential intersections in molecular plane, that out of the molecular plane and identification of potential intersections as JT-CI/RT-CI in different subsections. We like to mention that the states, for planar configuration, ($\varphi=0(2\pi),\pi$) are tagged with symmetry A´ and A″, while for non-planar configuration ($\varphi\neq0(2\pi),\pi$) are tagged by symmetry A.

### 3B. Potential Intersections in Molecular Plane:

The linear molecular system HCNO$^+$ has three consecutive low-lying doubly degenerate $^2\Pi$ states. Any deviation from linearity by shifting one end-atom from the molecular z-axis would lift this degeneracy leading to the generation of $^2$A´ and $^2$A″ states from each doubly degenerate $^2\Pi$ state. The lifting of degeneracy of two lowest $^2\Pi$ states, as a result of shifting the end atom H from the molecular axis, while maintaining co-planarity with the rest 3-atoms, is shown in Fig.1. We like to point out that the nature of variation of $1^2$A´ and $1^2$A″ (originated from linear $1^2\Pi$ state) with q$_H$ is qualitatively different from the nature of variation of $2^2$A´ and $2^2$A″ (originated from linear $2^2\Pi$ state) with q$_H$. Later we would generalize that this qualitative dissimilarity is related to the link of the original linear states respectively to partially occupied bonding ($\Pi$) and antibonding ($\Pi^*$) orbitals. It is to note that the symmetry of the planar molecule is chosen as C$_s$. For calculations of potential energy curves for such configurations, the following strategy has been taken : O-atom has been shifted by $q_O = 0.1$ Å from the molecular axis. With this fixed position for the O-atom, the H-atom is shifted in normal direction from the molecular axis by $q_H$, varying $q_H$ from -0.3 Å to + 0.3 Å , maintaining its position, co-planar with the plane defined by the rest of three atoms. The Fig.2A gives the corresponding geometry of the molecule. Adiabatic potential energy curves for the four low-lying states with $q_O$=0.1 Å and varying $q_H$ from -0.3 Å to +0.3 Å in perpendicular direction to the molecular C-N axis are given in panels 'a' and 'b' of Fig.3. The potential energy curves (PEC's) corresponding to $1^2A'$ and $1^2A''$, are seen to intersect each other for $q_H$ = -0.07 Å and $q_H$ = +0.15 Å. The electronic states $2^2A'$ and $2^2A''$, do not show any such intersection for such variation of $q_H$ in the molecular plane.

### 3C. Potential Intersection out of molecular plane:

Non-planar configurations of the tetra-atomic $HCNO^+$ are characterized by the dihedral angle φ, between planes (H, C, N) and (C, N, O), not equal to zero(2π) or π. Such configurations are obtained on rotation of the probe H-atom, while the dihedral angle between the planes (H , C , N) and (C , N , O) is not equal to zero(2π) or π. An example geometry is shown in Fig 2B. Now as the molecule attains a non-planar configuration, the states are tagged only by symmetry A. In Fig-4 panels 'a' and 'b' depict the potential energy curves for states $1^2A$ , $2^2A$ , $3^2A$ and $4^2A$, as $q_H$ is varied from -0.3 Å to +0.3 Å with $q_O$ fixed at 0.1 Å for a non-planar configuration corresponding to φ=$64^0$($116^0$). In panel 'c' of Fig.4 respectively the energy difference ($E_{3^2A}$ – $E_{4^2A}$ ) is plotted against $q_H$ for dihedral angle φ =0°, 64° and 116° with $q_O$ = 0.1 Å. This indicates clearly that there is a reversal of behaviour : states $1^2A$ and $2^2A$ intersect for φ = zero(2π) or π (in molecular plane) but not for θ≠ zero(2π) or π (out of molecular plane). Again, states $3^2A$ and $4^2A$ intersect for θ≠ zero(2π) or π (out of molecular plane ), but not for φ = zero(2π) or π (in molecular plane).

### 3D. Characterization of Potential Intersections

In this subsection we present the calculations of angular NACTs which aim to find the topological criteria of the potential intersections described in Figs 3a and 4a. For such calculations we chose atom O fixed at a distance $q_O = 0.1$ Å and the terminal H-atom as rotating with radius $q_H$ in a circular contour in a plane perpendicular to the axis formed by atoms C and N. The dihedral angle between the plane (H,C,N) and (C,N,O), referred as $\varphi$, varies from zero to $2\pi$. So, here, in general we designate the states with symmetry A as $1^2A$, $2^2A$, $3^2A$ and $4^2A$.

From Figs. 3 and 4, it is evident that the pair of states $3^2A$ and $4^2A$ are well separated from the pair of states $1^2A$ and $2^2A$. So after performing a 4-state ($1^2A$, $2^2A$, $3^2A$ and $4^2A$) SA-CASSCF calculation, one may do a 2-state ($1^2A$ and $2^2A$) MRCI to calculate NACT matrix elements $\tau_{12}$ (between $1^2A$ and $2^2A$). But for calculating NACT matrix elements $\tau_{34}$ (between $3^2A$ and $4^2A$) such protocol is not possible. So after a 4-state SA-CASSCF, we have performed a 4-state MRCI to calculate different NACT matrix elements such as $\tau_{12}$, $\tau_{34}$, $\tau_{13}$, $\tau_{14}$, $\tau_{23}$ and $\tau_{24}$ following DDR protocol[39]. However, as expected, NACT matrix elements $\tau_{13}$, $\tau_{14}$, $\tau_{23}$ and $\tau_{24}$ are very small, of the order of $10^{-3}$. The panels 'a', 'b' and 'c' of Fig.5 show respectively the variation of $\tau_{12}$, $\gamma_{12}$, $A_{11}$ ($A_{22}$) with $\varphi$, the angle defining the circularly closed contour for values of ($q_O$, $q_H$) as (0.1 Å, 0.1 Å) and (0.1 Å, 0.2 Å). For a clear understanding we refer to the potential intersections denoted by 'P' and 'Q' in Fig.3a. With atom 'O' shifted from molecular axis and clamped at $q_O$, the circular contour with radius $q_H$=0.1 Å encircles intersection 'P'. Corresponding calculations demonstrate a smooth variation of mixing angle $\gamma_{12}$ upto the value of topological phase $\alpha_{12} = 3.13$ (~$\pi$) at $\varphi=2\pi$, as indicated by the green line in Fig.5b. This indicates that intersection indicated by 'P' is a JT-CI. Now with the same position of atom 'O', the circular contour with radius as $q_H$=0.2Å will encircle both the intersections 'P' and 'Q'. Corresponding calculations, now, demonstrate a smooth variation of mixing angle $\gamma_{12}$ upto the value of topological phase $\alpha_{12} = 6.26$ (~$2\pi$) at $\varphi=2\pi$, as indicated by the violet line in Fig.5b. Keeping in mind about the contribution from the intersection 'P', one may conclude that intersection presented by 'Q' is also a JT-CI. The green line in Fig. 5c shows the evolution of $A_{11}$ (as well as $A_{22}$) from the value of +1 to that of -1 value of $D_{11}$ ($D_{22}$) for circular contour with radius $q_H = 0.1$ Å. The violet line in Fig.5c, shows that evolution from +1 to +1 value of $D_{11}$ ($D_{22}$) through the value of -1 for circular contour with radius $q_H$ =0.2 Å. These calculations are clear indication of existence of a pair of JT-CI's between the lowest pair of states, evolved from $1^2\Pi$ state, in the chosen CS. We point out that for the chosen CS, defined by $q_O=0.1$Å with $q_H=0.1$ Å and 0.2 Å, as has been shown in Fig.3a, these intersections lie in the molecular plane ($\varphi = $ zero($2\pi$), here) only and so we designate these intersecting states as $1^2A'$ and $1^2A''$.

Panels 'a', 'b' and 'c' of Fig.6 respectively show the variation of $\tau_{34}$, $\gamma_{34}$ and $A_{33}$ ($A_{44}$) with angle $\varphi$, for CS defined by $q_O =0.1$Å and $q_H = 0.1$ Å. The observation of $\alpha_{34}$ as ~$2\pi$ at $\varphi=2\pi$ and evolution of $A_{33}$ ($A_{44}$) from +1 to +1 value of $D_{33}$ ($D_{44}$) through the value of -1 indicate the existence of a RT type intersection between states $3^2A$ and $4^2A$. From Fig.3b (corresponding to $\varphi = $ zero($2\pi$) ) and Fig.4b (corresponding to $\varphi = 64^0(116^0)$ ), it may be noted clearly that the potential intersection between states evolved from $2^2\Pi$ state takes place only for some non-planar molecular configuration corresponding to $\varphi=64^0(116^0)$ and not in molecular plane ($\varphi = $ zero($2\pi$) or $\pi$). Variation of $\tau_{34}$ and $\gamma_{34}$ for lower value of $q_H < 0.1$ Å are also shown in Fig.6 to substantiate that there is no JT-CI present in the CS defined by $q_O =0.1$Å and $q_H <0.1$ Å, corresponding to $\varphi=64^0$ ($116^0$). Thus, here $\alpha_{34}$ ~$2\pi$ is due to the presence of RT-CI only. We conclude that the original linear $2^2\Pi$ state, evolves into the states $3^2A$ and $4^2A$ and these two states ($3^2A$ and $4^2A$) undergo a RT-PI out of molecular plane.

**4. Orbital Analysis of Plane of Appearance of Potential Intersections:**

Present work shows that as bending is introduced in the linear molecular system HCNO$^+$ with terminal atoms H and O shifted from the molecular axis by values as designated by $q_H$ and $q_O$ respectively, it shows a pair of JT-CI's (involving $1^2A'$ and $1^2A''$) in the molecular plane and one RT-PI involving states $3^2A$ and $4^2A$ out of the molecular plane. The states $1^2A'$ and $1^2A''$ evolve from linear $1^2\Pi$ state, while the states $3^2A$ and $4^2A$ evolve from linear $2^2\Pi$ state, with onset of bending. We want to reiterate here, that earlier calculations have shown that two states evolved from X$^2\Pi$ state, on introduction of bending, show potential intersections in molecular plane for C$_2$H$_2^+$ [17], HBNH$^+$ [28] and in some non-planar configuration for N$_2$H$_2^+$ [28], HCNH [25]. In the present section we want to explore qualitatively what is common in the X$^2\Pi$ states of C$_2$H$_2^+$, HBNH$^+$ with $1^2\Pi$ state of HCNO$^+$ and similarly among the X$^2\Pi$ states of N$_2$H$_2^+$, HCNH and $2^2\Pi$ state of HCNO$^+$.

As stated in Section 3, we have chosen 15 electrons as valence electrons out of 21 electrons in HCNO$^+$. The active space in SA-CASSCF calculation has been constituted by distributing these 15 electrons in 12 orbitals. SA-CASSCF calculation has been followed by MRCI calculations. In state-averaged calculation, contribution to CI vector loses its significance; still it remains true that self-consistent-field (SCF) determinant would be the highest contributing determinant. We give below the electron distribution for the SCF determinant corresponding to linear $1^2\Pi$ state.

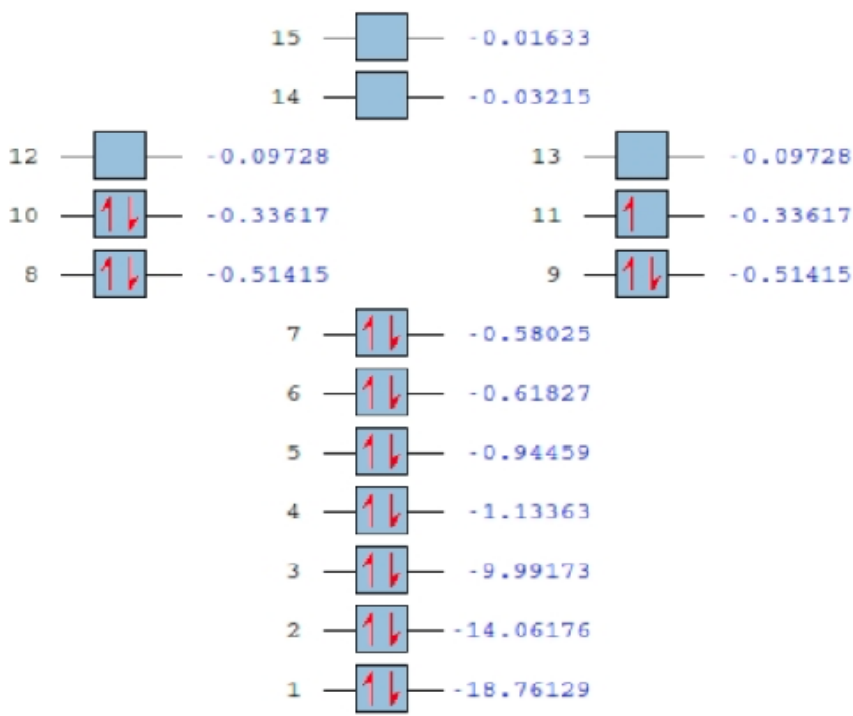

The partial occupancy of the $\Pi$-orbitals (10 & 11) may be noted. One can check that the electron distribution for the qualitatively speculated most contributing determinant for the linear $2^2\Pi$ state differs from that for $1^2\Pi$ state by the fact that for $2^2\Pi$ state, the partially occupied orbitals are the $\Pi^*$-orbitals (12 & 13) only. In Fig.7 we present the variation of energy of $\Pi/\Pi^*$ orbital, thus associated to $1^2\Pi$ / $2^2\Pi$ states of HCNO$^+$, with variation of $q_H$ from +0.3 to -0.3. It is noteworthy that for planar configuration of molecule, the variation for the case of $\Pi$ orbital is similar to the variation of potential energy for $1^2Á$ and $1^2Á''$ (Fig.3a) for HCNO$^+$, evolved from the $1^2\Pi$ state, as $q_H$ is varied similarly. Again for a non-planar molecular configuration variation of energy for the case of $\Pi^*$ orbital, it is similar to the variation of potential energy for $3^2A$ and $4\ ^2A$ (Fig.4b), evolved from $2^2\Pi$ for similar variation of $q_H$. These findings give us a motivation for looking at these $\Pi/\Pi^*$ orbitals. Now we are in a position to check whether this idea may be extended to other similar molecules ($C_2H_2^+$, HCNH, $C_2H_2^+$ and $HBNH^+$) studied earlier. It has been checked from calculation, that for the plausible most contributing determinants, the partially occupied $\Pi$-orbitals are bonding for the concerned $X^2\Pi$ states of $C_2H_2^+$, $HBNH^+$ and anti-bonding for concerned $X^2\Pi$ states of $N_2H_2^+$, HCNH. We advocate that thus the $X^2\Pi$ states of $C_2H_2^+$, HCNH, $C_2H_2^+$, $HBNH^+$ as well as $HCNO^+$ may be classified as related to partially occupied $\Pi/\Pi^*$ orbitals.

In Fig.8 we have depicted the partially filled $\Pi$ and $\Pi^*$ orbitals involved in non-planar configuration of $HCNO^+$, corresponding to dihedral angle between the planes (H,C,N) and (C,N,O) as $116^0(64^0)$. A careful look into the figure reveals that for this non-planar configuration, the probe H-atom is interacting more promisingly with the $\Pi^*$ orbital, in comparison to the $\Pi$ orbital. This qualitatively justifies the appearance of potential intersection between the states $3^2A$ and $4^2A$ (evolved from linear $2^2\Pi$ state, related to $\Pi^*$-MO by origin) in the corresponding non-planar configuration. We advocate that if the original $X^2\Pi$ state is related to $\Pi^*$ orbital, then the intersection between two evolved states appear out of molecular plane. Similarly for planar molecular configuration, the probe H-atom interacts better with the $\Pi$-orbital. This, again, qualitatively justifies the appearance of potential intersection between $1^2A'$ and $1^2A''$ (evolved from linear $1^2\Pi$ state, related to $\Pi$-MO by origin), in molecular plane.

Table I shows the plane of appearance of potential intersections (JT-CI's and RT-PI's) along with the character of $\Pi$-MO's involved in the concerned original linear states of $C_2H_2^+$ [17], $HC_2O$ [27], HCNH[25], $N_2H_2^+$ [28] and $HBNH^+$ [28]. An analysis of the result accumulated in the Table I leads to the generalization that for a slightly bent linear tetra-atomic, two states evolved from a linear $X^2\Pi$ state, related to a $\Pi$-MO by origin, shows a potential intersection in molecular plane, while states evolved from a linear $X^2\Pi$ state, related to a $\Pi^*$-MO by origin, show a potential intersection out of molecular plane.

## 5. Conclusion:

It has already been pointed out that JT-CI between two lowest states in bent configuration ($1^2A'$ and $1^2A''$, originated from collinear $X^2\Pi$ state) appear in the molecular plane for $C_2H_2^+$ [17] and $HBNH^+$ [28] while out-of-the molecular plane for $N_2H_2^+$ [28] and $HCNH^+$ [25]. We reiterate - this indicates clearly that such phenomena has no relation with the existence/non-existence of a plane of reflection in the molecule, as may appear at first place. Our present study along with the result of the earlier ones[25,28], as accumulated in Table I, advocate that an orbital analysis of the involved states shows that bonding / antibonding nature of the highest partially filled Π-MO's in the qualitatively most contributing determinant, may be used to predict the plane of appearance of such possible intersections.

Here, we like to highlight the importance of finding whether the intersection between two states appear in molecular plane or not. To appreciate this we may take the case[26] of dissociation of HCNH molecule to the formation of HNC: while HCNH is in its $1^2\Sigma^+$ state, which is dissociative in nature, then only C-H bond will dissociate leading to formation of HNC. Consideration of non adiabatic interaction among three low lying states $1^2\Pi$, $1^2\Sigma^+$, $2^2\Sigma^+$ of HCNH molecule, in its bent configuration, reveal[26] that the C-H bond will dissociate only if the two $^2\Sigma^+$ states interact with each other. Das and Mukhopadhyay[26] have investigated that there occurs a same-symmetry JT-CI between two $^2\Sigma^+$ states, and then they evolve as $2^2A'$ and $3^2A'$ only out of the molecular plane. This implies the importance of the detection of the plane of appearance of the intersection between the two states. However, our present project deals with the states evolved from $X^2\Pi$ state only.

**Acknowledgement** : R.M. acknowledges UGC, India for a fellowship. D.M. acknowledges BRNS (DAE, India) Project (No.2009/37/42/BRNS) for computational facility.

**Figure and table :**

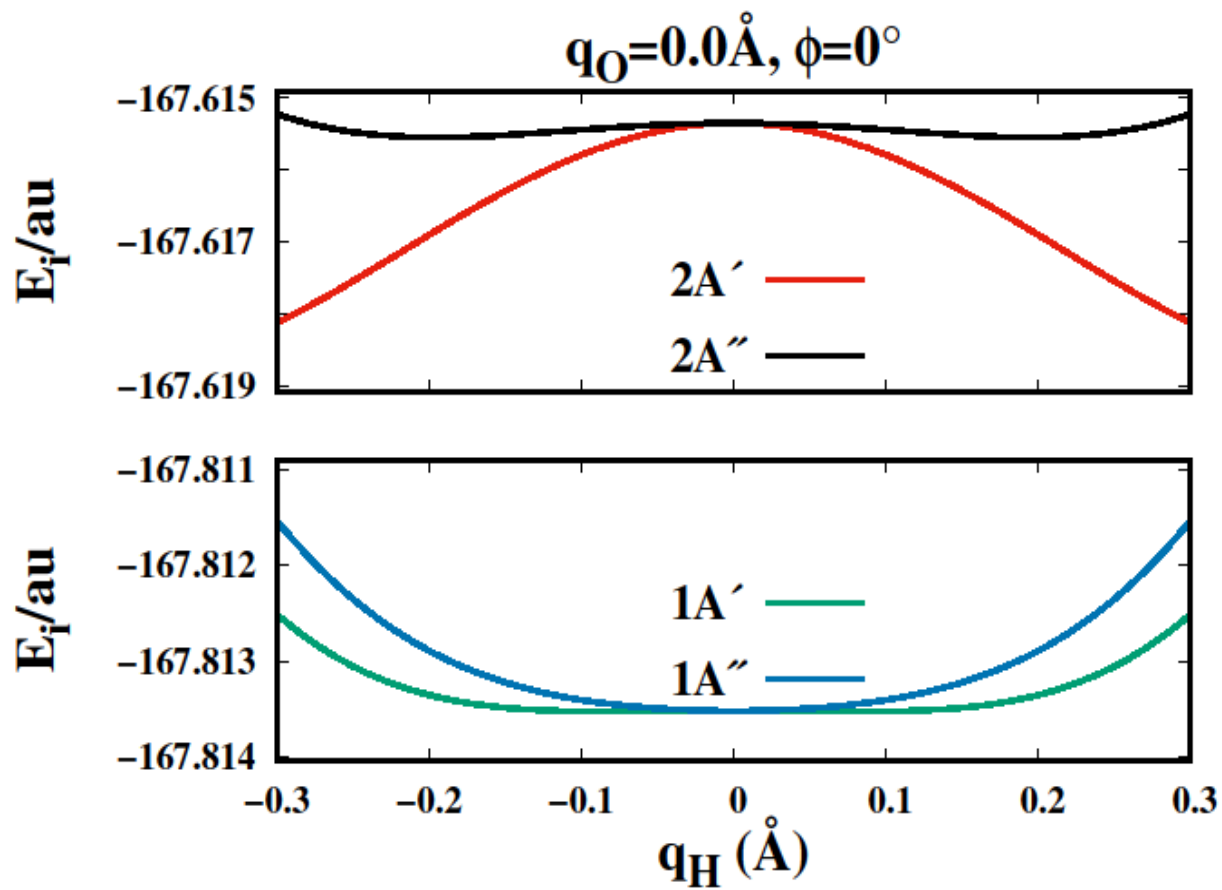

**Figure 1.** Variation of energy of 4 low-lying states with variation of $q_H$ from -0.3Å to +0.3Å in molecular plane (i.e. $\varphi=0$ ), without shifting the O-atom form molecular axis (i.e. $q_O=0.0$). In notation of states, spin-multiplicity is dropped for simplicity in presentation.

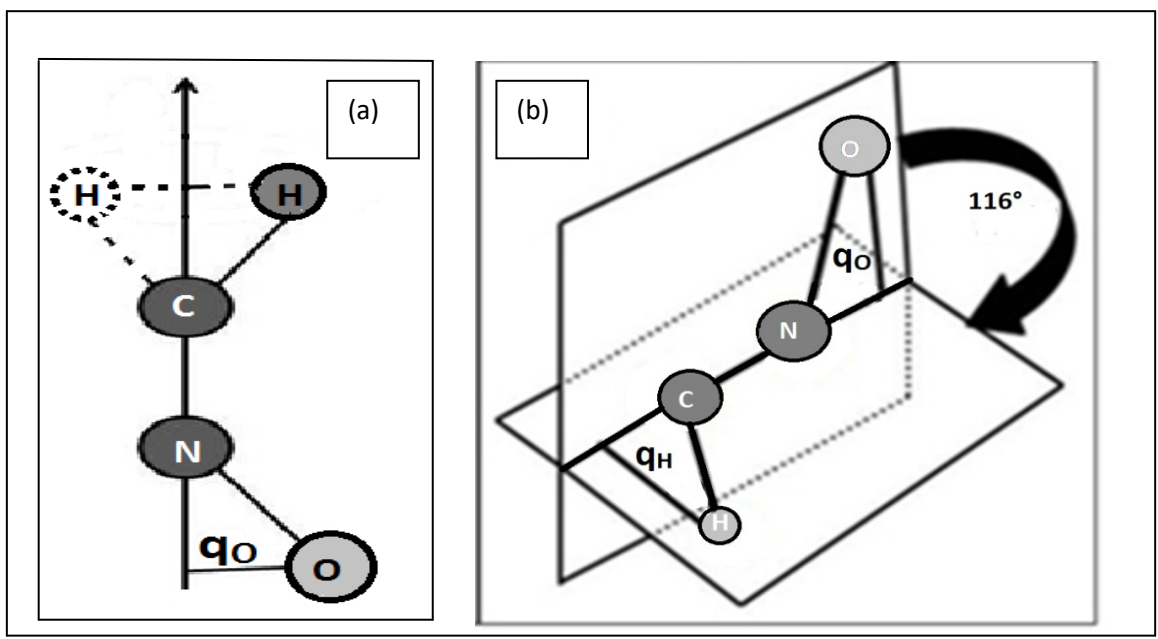

**Figure 2.** Panel (a): Arrangement of four atoms of HCNO$^+$ molecule: O-atom shifted from the molecular axis with certain fixed q value ($q_O$) and the H-atom moves along the normal to the molecular axis with varying q values ($q_H$). Panel(b):The non planar geometry of the molecule where the planes (C,N,O) and (H,C,N) make a dihedral angle between them as 116°.

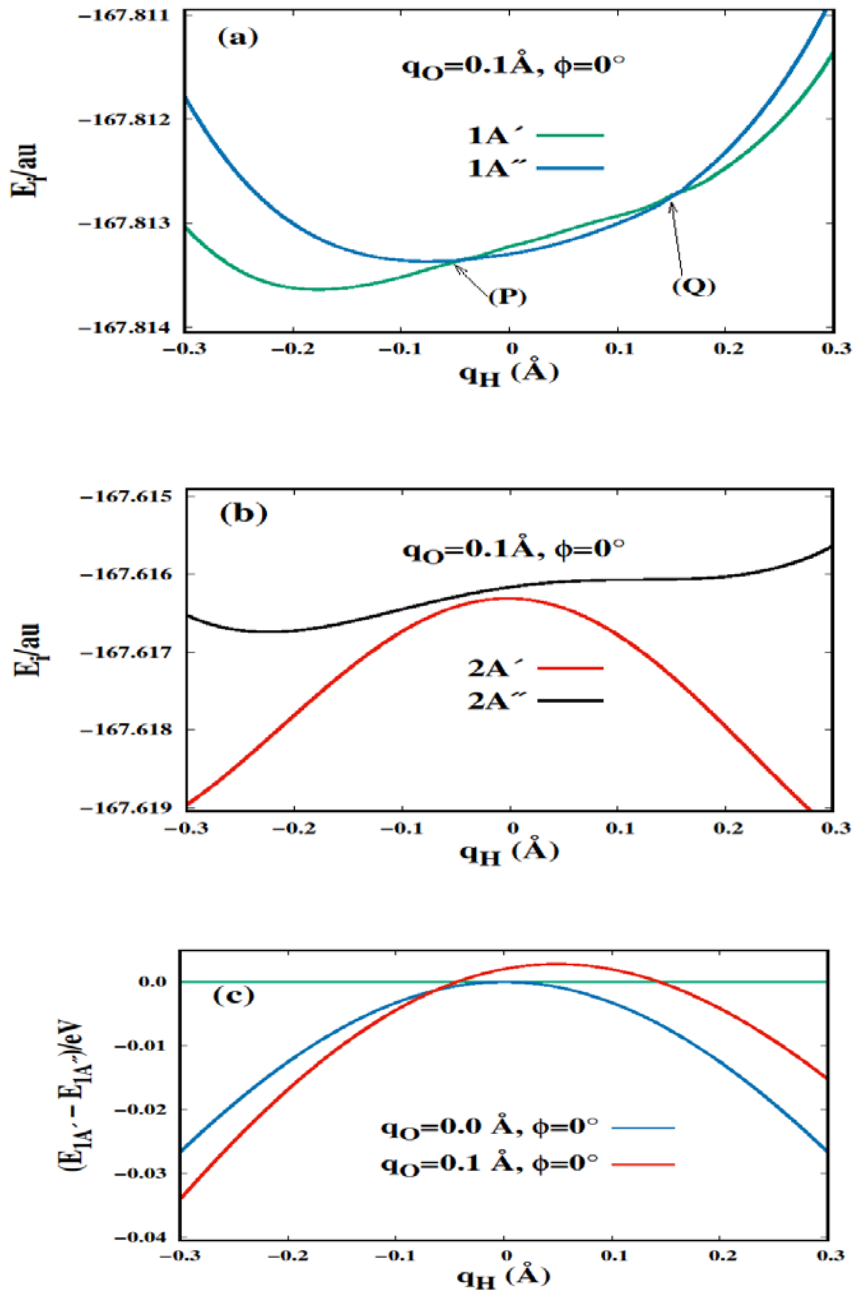

**Figure 3.** Panel (a): Variation of the energy of two low-lying electronic states $1^2\acute{A}$ and $1^2A''$ with $q_H$, varying from -0.3 Å to +0.3 Å for ($q_O$=0.1 Å, $\varphi$=0 ).
Panel (b): Variation of the energy of two states $2^2\acute{A}$ and $2^2A''$ with $q_H$, varying from -0.3 Å to +0.3 Å for ($q_O$=0.1 Å, $\varphi$=0 ).
Panel (c ): Variation of energy difference of two states $1^2\acute{A}$ and $1^2A''$ with $q_H$ for ($q_O$=0.0 Å, $\varphi$=0 ) and for ($q_O$=0.1 Å, $\varphi$=0 ). In notation of states, spin-multiplicity is dropped for simplicity in presentation.

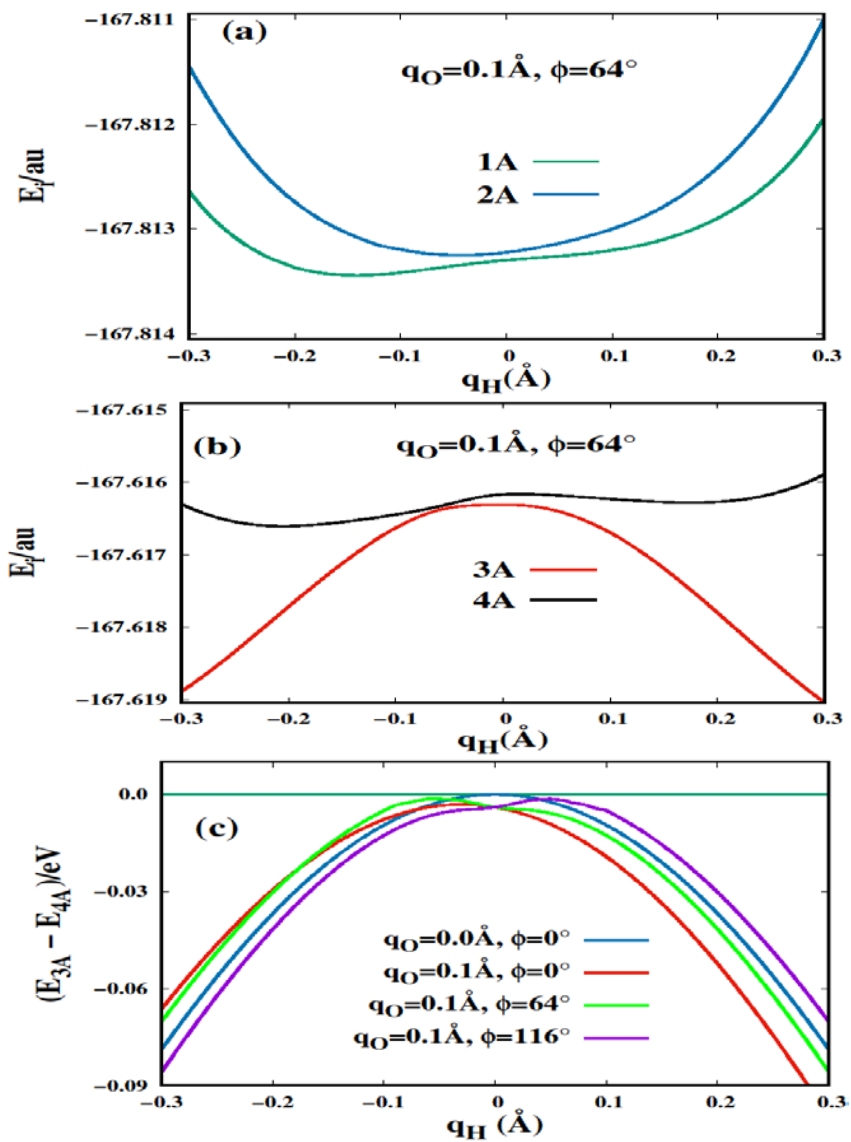

**Figure 4.** Panel (a): Variation of the energy of two low-lying electronic states $1^2A$ and $2^2A$ with $q_H$, varying from -0.3 Å to +0.3 Å for ($q_O$=0.1 Å, $\varphi$=64°)
Panel (b): Variation of the energy of two states $3^2A$ and $4^2A$ with $q_H$, varying from -0.3 Å to +0.3 Å for ($q_O$=0.1 Å, $\varphi$=64°).
Panel (c): Variation of energy difference of two states $3^2A$ and $4^2A$ with $q_H$ for ($q_O$=0.0 Å, $\varphi$=0), ($q_O$=0.1 Å, $\varphi$=0), ($q_O$=0.1 Å, $\varphi$=64) and ($q_O$=0.1 Å, $\varphi$=116°).
In notation of states, spin-multiplicity is dropped for simplicity in presentation.

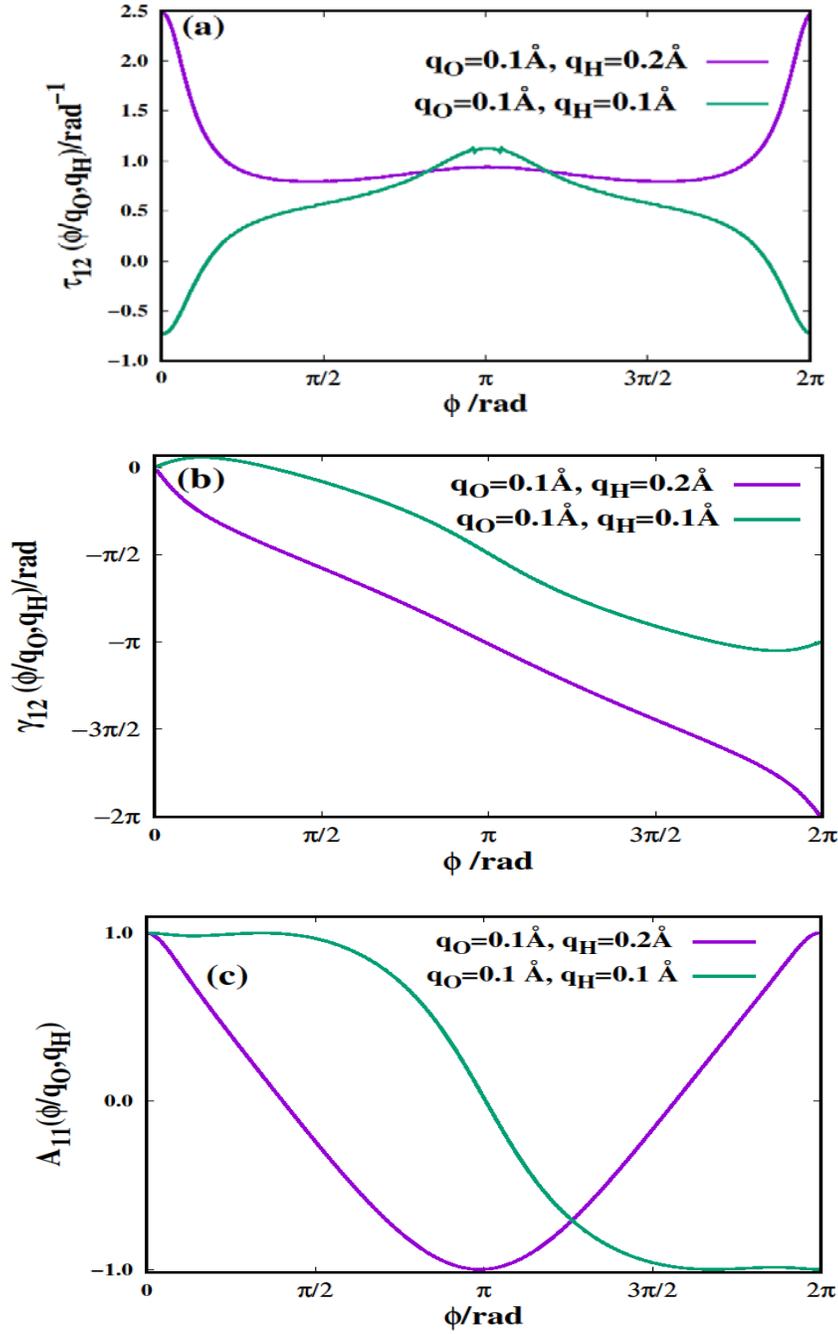

**Figure 5.** Panels (a), (b) and (c) represent respectively the variation of the angular NACT $\tau_{12}(\varphi|q_H,q_O)$, the ADT angle $\gamma_{12}(\varphi|q_H,q_O)$ and the diagonal matrix element $A_{11}(\varphi|q_H,q_O)$ of the (4X4) ADT matrix as a function of $\varphi$ for ($q_O$=0.1 Å, $q_H$=0.1 Å) as well as ($q_O$=0.1 Å, $q_H$=0.2 Å). In notation of states, spin-multiplicity is dropped for simplicity in present

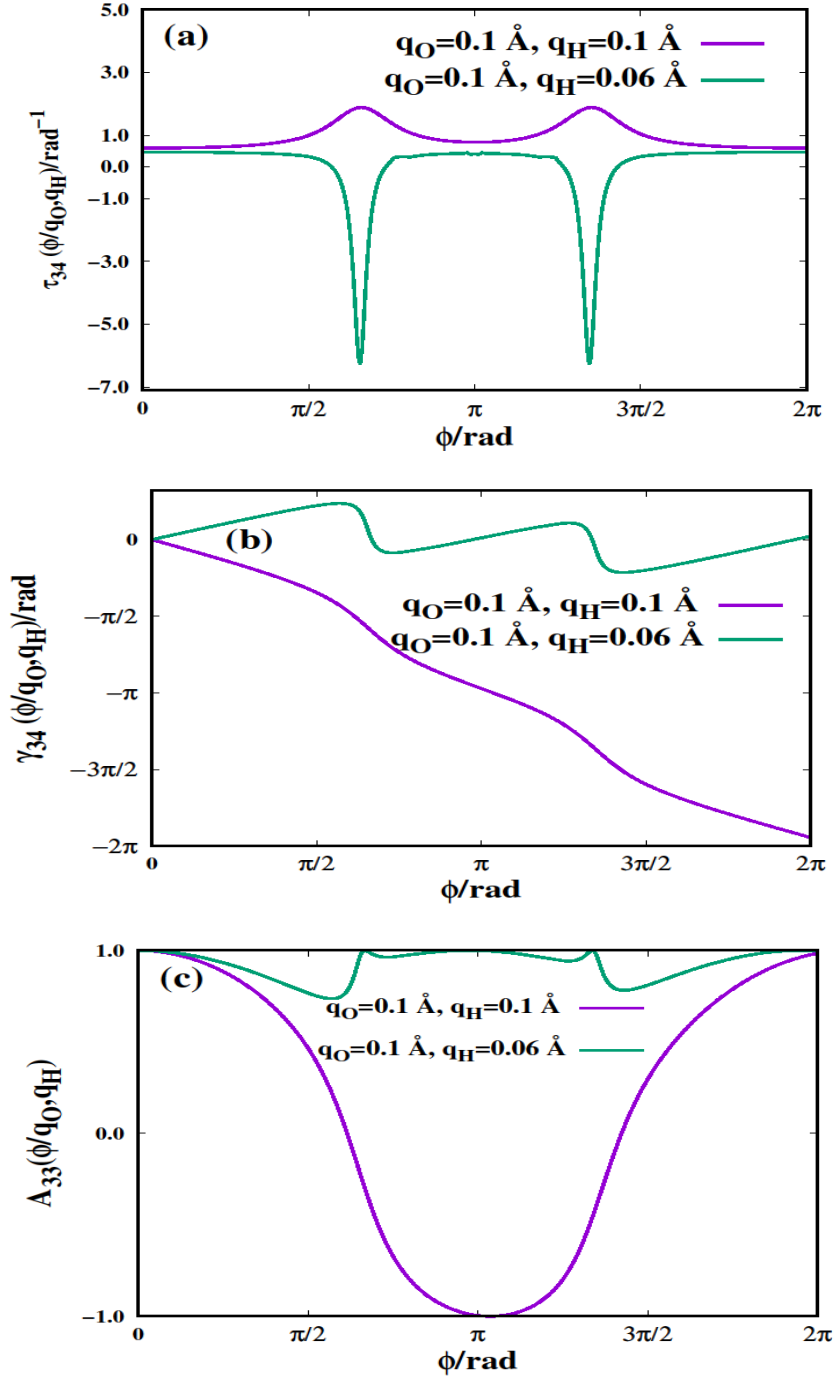

**Figure 6.** Panels (a), (b) and (c) represent respectively the variation of the angular NACT $\tau_{34}(\varphi|q_H,q_O)$, the ADT angle $\gamma_{34}(\varphi|q_H,q_O)$ and the diagonal matrix element $A_{33}(\varphi|q_H,q_O)$ of the (4X4) ADT matrix as a function of $\varphi$ for ($q_O$=0.1 Å, $q_H$=0.1 Å ) as well as ($q_O$=0.1 Å, $q_H$=0.06 Å ). In notation of states, spin-multiplicity is dropped for simplicity in presentation.

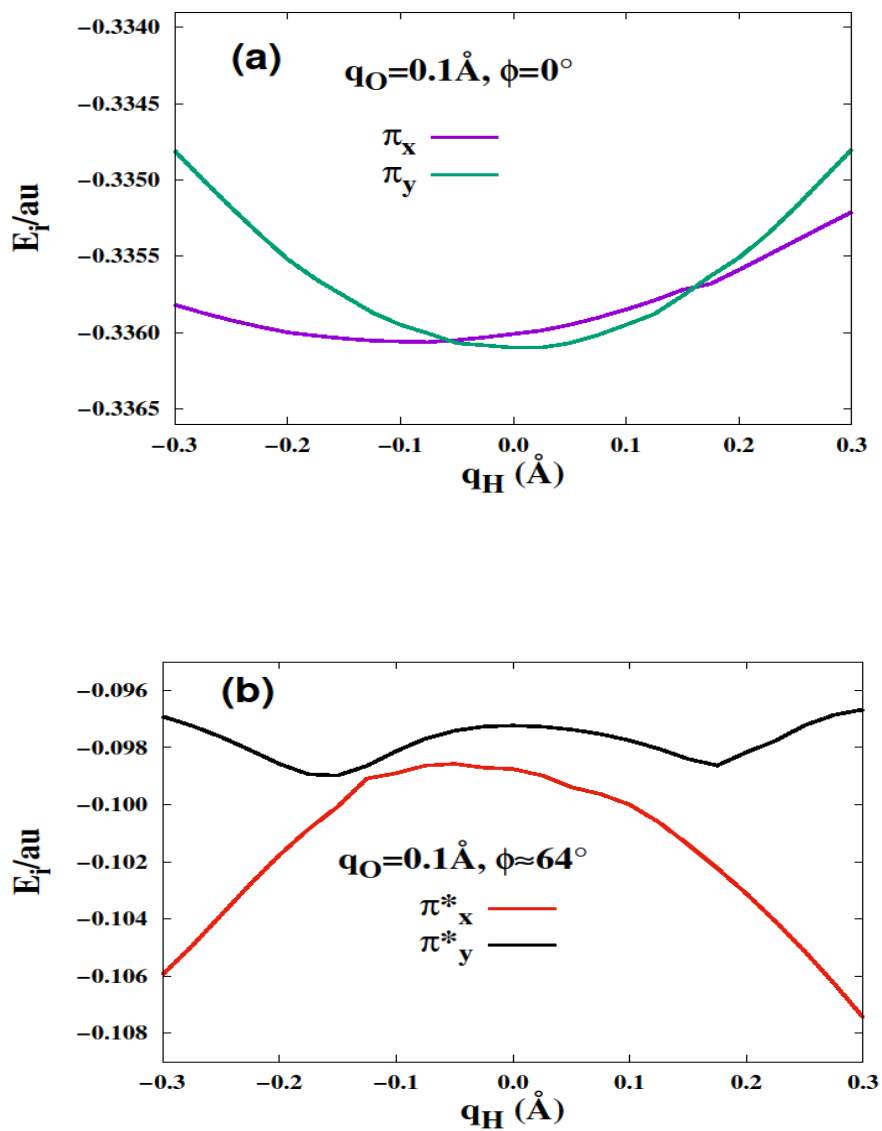

**Figure 7.** Panel (a) gives variation of energy of partially filled $\Pi$ molecular orbital (M.O.) related (please see text) to $1^2\Pi$ state of $HCNO^+$ with variation of $q_H$ for $\varphi=0^0(2\pi)$. Panel (b) gives variation of energy of partially filled $\Pi^*$-M.O. related to $2^2\Pi$ state of $HCNO^+$ with variation of $q_H$ for $\varphi\approx 64^0$.

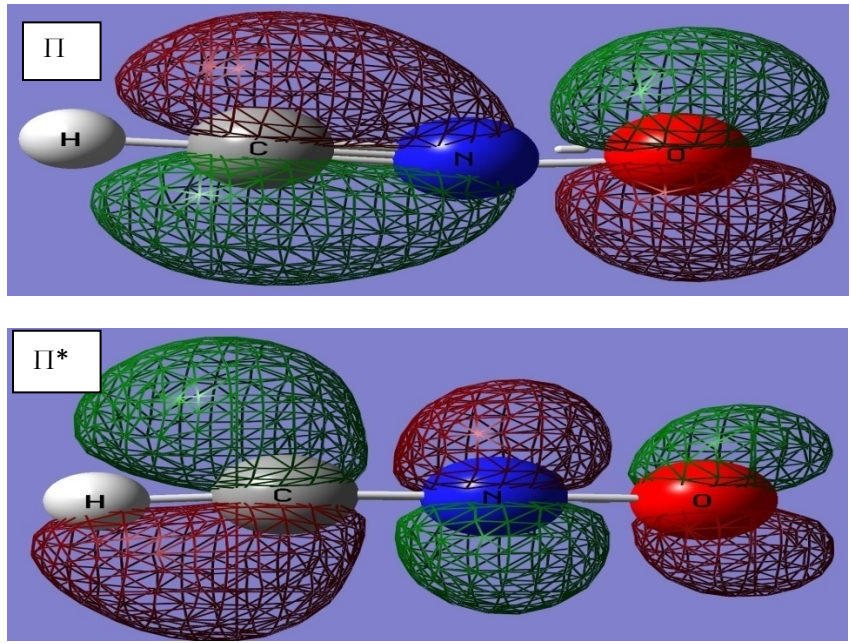

**Figure 8**. Π and Π* Molecular Orbital of HCNO⁺ molecule in non-planar configuration when φ =116°(64⁰).

**Table I.** Summary of the result so far available with the tetra atomic molecular systems and a correlation between the nature of partially filled Π-MO's in the most contributing determinant of the linear $X^2\Pi$ state and the plane of appearance of the potential intersections observed.

| System studied | Electronic states involved | MO's Related to original linear $X^2\Pi$ state | Plane of appearance of JT/RT intersections |
|---|---|---|---|
| $C_2H_2^+$ | 1A′ and 1A″ | Π | In Molecular plane |
| $N_2H_2^+$ | 1A and 2A | Π* | Out of molecular plane |
| $HBNH^+$ | 1A′ and 1A″ | Π | In molecular plane |
| HCNH | 1A and 2A | Π* | Out of molecular plane |
| $HCNO^+$ | 1A′ and 1A″ | Π | In molecular plane |
| $HCNO^+$ | 3A and 4A | Π* | Out of molecular plane |